# Excitation of spin dynamics by spin-polarized current in vortex state magnetic disks

B. A. Ivanov [1] and C. E. Zaspel [2]

[1] Institute of Magnetism, 03142 Kiev, Ukraine; [2] University of Montana-Western, Dillon, MT 59725, USA

A spin-polarized current with the polarization perpendicular to the plane of a vortex-state disk results in renormalization of the effective damping for a given magnetization mode, and the effective damping becomes zero if the current exceeds a threshold value, $I_c$. The lowest $I_c$ corresponds to the lowest frequency, $\omega_{GM}$ vortex gyroscopic mode. For $I > I_c$, the dynamic magnetization state is characterized by precession of the vortex around the dot center with non-small amplitude and higher frequency, $\omega > \omega_{GM}$.

The injection of a spin current in magnetic nanostructures can result in interesting new effects having potential applications in spintronics. For example, it was first theoretically predicted by Slonczewski [1] and Berger [2] that a dc spin polarized current in a uniformly magnetized thin film renormalizes the spin wave damping and above a critical value of the spin current, $I_c$ can result in negative damping and generation of spin waves. For a more complete theoretical description of the nature of the modes excited by a spin-polarized current it is necessary to take nonlinear effects into account. In the thin film it was shown [3] that nonlinearity from the four-wave interaction will limit the growth of the excitation. Moreover, it was predicted [4] that the stable structure will be a self-localized spin wave bullet. In nanomagnets such as thin disks the ground state structure is vortex-like providing an ideal system to study spin-transfer effects in confined nonuniform states. For these systems it was recently shown [5] that a spin-polarized current with the polarization in the vortex plane results in a displacement of the vortex core owing to the lower symmetry introduced by the current. More recently for the simple model of the easy-plane magnet it was theoretically predicted that the vortex core polarity can be switched [6] by a spin current perpendicular to the vortex plane, however, some of the important magnetostatic effects were neglected. This particular situation has direct application to information storage since the vortex core polarization as well as the vortex chirality can potentially each store one bit of information. In this letter the nonlinear dynamic effects from a dc spin polarized current are investigated including dominant magnetostatic effects from both surface and volume magnetostatic charges.

Since the dynamic effects from a spin polarized current are theoretically investigated here, we begin with a short introduction to the spectrum of vortex-state disks. The lowest frequency (sub GHz) excitation corresponds to gyrotropic oscillation of the vortex core as a result of an initial displacement induced by an in-plane magnetic pulse [7,8]. There are also higher frequency (GHz) modes excited [9-11] and their structure depends on the symmetry of the initial pulse. In particular, the radially symmetric mode can be excited by an out-of-plane pulse.

To determine how a current will affect these excitations, the Landau-Lifshitz equations are analyzed including effects from the spin-polarized current and dissipation.

For the particular case were the polarization direction, $\hat{p}$ is perpendicular to the vortex plane, and the cylinder thickness, $L$ is small compared to the radius, $R$ the magnetization can be assumed to be uniform in the perpendicular ($z$-direction) and the Slonczewski form [1] for the spin torque is applicable,

$$\vec{\tau} = \sigma I M_s \left[ \vec{m} \times (\vec{m} \times \hat{p}) \right], \qquad (1)$$

where $I$ is the current and $\sigma = \varepsilon g \mu_B / 2 e M_s L A$ where $\varepsilon$ is the spin-polarization efficiency, $e$ is the magnitude of the electric charge, $g$ is the Lande factor, $\mu_B$ is the Bohr magneton and $A = \pi R_c^2$ is the area of the circular nanocontact, where $R_c = R$ in the following. Also this is a function of the unit magnetization vector, $\vec{m} = \vec{M}/M_s$ where $M_s$ is the saturation magnetization, and this can be expressed ($\vec{m} = \sin\theta\cos\varphi \hat{r} + \sin\theta\sin\varphi \hat{\chi} + \cos\theta \hat{z}$) in terms of polar and azimuthal angles, $\theta$ and $\varphi$, respectively.

Taking into account dissipation and the Slonczewski torque, the equations of motion for the magnetization have the following form

$$\frac{M_s}{\gamma} \sin\theta \frac{\partial \varphi}{\partial t} = \frac{\delta W}{\delta \theta} + \frac{\delta Q}{\delta \dot{\theta}}, \qquad (2)$$

$$-\frac{M_s}{\gamma} \sin\theta \frac{\partial \theta}{\partial t} = \frac{\delta W}{\delta \varphi} + \frac{\delta Q}{\delta \dot{\varphi}} - \frac{\sigma I M_s \sin^2\theta}{\gamma}, \qquad (3)$$

for the particular case when $\hat{z}$ is the polarization direction. Here $Q$ is the dissipation function

$$Q_r = \frac{\alpha M_s}{2\gamma} \int \left( \dot{\theta}^2 + \sin^2\theta \dot{\varphi}^2 \right) d\vec{r}, \qquad (4)$$

which results in Gilbert damping.

First it is shown that there will be a critical current where damping will be balanced by the Slonczewski torque. This is easily done through an energy balance estimate, by multiplication of (2) by $\dot{\varphi}$, multiplication of (3) by $\dot{\theta}$ and integration over the disk volume to get the time derivative of the energy

$$\frac{dW}{dt} = -\frac{\alpha L M_s}{\gamma} \int \left(\dot\theta^2 + \sin^2\theta \dot\varphi^2\right) d^2x +$$

$$+ \frac{\sigma I M_s L}{\gamma} \int \sin^2\theta \dot\varphi d^2x. \qquad (5)$$

Since it is assumed that the disk is thin, the magnetization is assumed to be constant in the $z$ direction. As expected, dissipation will result in energy loss, but the second term can be positive so there can be a critical current, $I_c$ where the energy dissipation is zero, and for $I > I_c$ the mode becomes unstable and its amplitude will increase.

To proceed further for an arbitrary mode, use the usual ansatz [9] $\theta = \theta_0 + \vartheta$, $\varphi = \chi + \pi/2 + \mu/\sin\theta_0$, where $\theta_0$ describes the vortex profile, $\vartheta_{m,n}(r,\chi,t) = f_{m,n}(r)\cos(m\chi - \omega t)$ and $\mu_{m,n}(r,\chi,t) = g_{m,n}(r)\sin(m\chi - \omega t)$ determine the magnon modes with an azimuthal characteristic number, $m$ and radial characteristic number, $n$. Here $r, \chi$ are polar coordinates in a dot plane, the forms of functions $f_{m,n}(r)$, $g_{m,n}(r)$ are well-known [11,12]. Then the condition $dW/dt = 0$, which is the definition of the critical current, $I_c$ gives

$$\alpha\omega \int (f^2 + g^2) r dr = \sigma I_c \int fg \cos\theta_0 r dr. \qquad (6)$$

To find the value of $I_c$, we need definite forms of the functions $f_{m,n}(r)$, $g_{m,n}(r)$, but some general remarks can be made. First, it is noticed that for small $\alpha \ll 1$ the mode damping decrement from Gilbert dissipation can be written as $\Gamma = Q/\delta E$, where $\delta E$ is the mode energy, calculated using the particular mode with $\alpha = 0$. The simple calculations show that, the expression for $\Gamma$ contains the same combinations of integrals as in (6), and the threshold value can be expressed through the values of $\Gamma$ and the mode frequency $\omega$,

$$I_c = \frac{\Gamma}{\sigma\omega}. \qquad (7)$$

Therefore, in the following the critical current is proportional to the ratio $\Gamma/\omega$. The values of $\omega$

for different modes are well known, and $\Gamma$ for some modes are determined experimentally [13], and the lowest $\Gamma/\omega$ corresponds to vortex gyrotropic mode [14]. One can expect that as the current increases the smallest critical current will correspond to amplification of the sub-GHz gyrotropic mode. As we will see below, the value can be as small as a few tens of microamps; whereas, for other modes (including the $m = 0$ mode corresponding to core size oscillations, considered in [6]) our calculations gives higher values in the milliamp range.

Since the gyrotropic mode has the lowest value of $I_c$, let us concentrate on quantitative analysis of this mode, determining the current dependence of the mode amplitude. The gyrotropic mode in the nonlinear regime can by modeled by the Thiele equation [15], and the easiest way to derive this equation including the Slonczewski torque is multiplication of (2) by $\nabla\varphi$, (3) by $\nabla\theta$, add and integrate over the disk volume. Then use of the vortex ansatz, $\vec{m} = \vec{m}_0(\vec{r} - \vec{X})$, where $\vec{m}_0(\vec{r})$ is the magnetization of the static vortex at the origin, gives the Thiele equation

$$\vec{G} \times \frac{d\vec{X}}{dt} - \frac{\partial W}{\partial \vec{X}} - \eta \frac{d\vec{X}}{dt} + \vec{F}_{SPC} = 0, \tag{8}$$

where $\vec{G} = -\hat{z} 2\pi M_s L / \gamma$ is the gyrovector [15,16], the viscosity coefficient is

$$\eta = \frac{M_s L}{\gamma} \int \alpha\left((\nabla\theta)^2 + \sin^2\theta(\nabla\varphi)^2\right) d^2x, \tag{9}$$

and the spin polarized current (SPC) force is

$$\vec{F}_{SPC} = \sigma I \frac{M_s L}{\gamma} \int \nabla\varphi \sin^2\theta d^2x. \tag{10}$$

From previous work the precession frequency can be obtained from the first two terms of (8) using $V = \omega X$ and the quadratic dependence of the magnetostatic energy, $W = \kappa X^2/2$ to get simply $\omega = \kappa G$. Also from (8), the critical current can be obtained by equating the damping force to the SPC force, but to accomplish this it is next necessary to evaluate the integrals in (8) and calculate the magnetostatic energy for the displaced vortex.

For the evaluation of (9) and (10) it is a good approximation to neglect the contribution from the core structure and evaluate the radial integrals in the range $l_0 \leq r \leq R$ where $l_0 = \sqrt{A/4\pi M_s^2}$ is the exchange length, $A$ is the exchange constant. For permalloy the core radius is the order of $l_0$ which is about 5 nm. In this case $\theta = \pi/2$ and the viscosity coefficient is given by

$$\eta = \frac{\alpha \pi M_s L}{\gamma} \int \left[ \left(\frac{d\theta}{dr}\right)^2 + \frac{\sin^2 \theta}{r^2} \right] r dr, \tag{11}$$

or, in main logarithmic approximation, $\eta = (\pi \alpha M_s / \gamma) \ln(R/l_0)$. Using equation (8), the value of $\eta$ can be connected with damping and frequency of gyrotropic mode, $\Gamma/\omega = \eta/G$, and compared with the values of $\Gamma$ found by Guslienko [14].

When the centered ($\vec{X} = 0$) vortex structure is used to evaluate (10) one obtains zero so it is necessary to assume a form for the displaced vortex structure. Previous work indicates that boundary conditions are approximately fixed at the disk edge resulting in no net edge magnetostatic charge [7,8]. For this reason it is useful to use the vortex-image vortex ansatz

$$\varphi(x, y) = \tan^{-1}\left(\frac{y}{x-a}\right) + \tan^{-1}\left(\frac{y}{x - R^2/a}\right) + \frac{\pi}{2}, \tag{12}$$

where $R$ is the disk radius and $a$ is the displacement of the vortex center on the $x$ axis, and the image vortex is outside the disk at $R^2/a$ also on the $x$ axis. The more general ansatz [8] for an arbitrary displacement is slightly more complicated, but this simpler form is sufficient. Using (12) it is possible to evaluate (10) exactly to obtain

$$\vec{F}_{SPC} = 2\pi L\sigma IM_s \left(\hat{z} \times \vec{X}\right) \tag{13}$$

for an arbitrary displacement. Thus, for circular vortex motion with constant frequency the SPC force is antiparallel to the friction force, and both are perpendicular to vortex displacement. In contrast, both the gyroforce and restoring force, for any $W = W(|\vec{X}|)$, have a radial direction. Then, the condition $\omega a = dW/da$ gives the frequency for amplitude $a = |\vec{X}|$, and the condition $2\pi L\sigma IM_s = \omega\eta$ determine the value of current resulting in this motion.

The remaining force in (8) is from the magnetostatic potential originating from the volume magnetostatic charge density, $\nabla \cdot \vec{M}$ which will be obtained using the magnetization expressed by (12). This is the critical part where nonlinearities become important owing to the displacement of the vortex core. For the region outside the core the normalized magnetization is given by $m_x = M_s$ and $m_y = M_s \sin\varphi$ with $m_z = 0$. Next these components of the magnetization are expanded in series on $a/R$,

$$m_x = m_{x0} + m_{x1}\frac{a}{R} + m_{x2}\frac{a^2}{R^2} + \cdots \quad (14)$$

For the consideration of nonlinear effects, third and fourth order terms should be used. For evaluation of integrals it is convenient to use a cylindrical $(r,\chi,z)$ coordinate system. We present here few lowest terms only, $m_{x0} = -\sin\chi$, $m_{y0} = \cos\chi$, $m_{x1} = -[(R^2 - r^2)/2Rr]\sin 2\chi$, $m_{x1} = -[(R^2 - r^2)/Rr]\sin^2\chi$, and $m_{y2} = (2r^2R^2)^{-1}(r^4 + 2r^2R^2 - 3R^4)\sin^2\chi\cos\chi$. The term $m_{x2}$ have the similar structure to $m_{y2}$, other terms are easily obtained but are too long to include here.

The main contribution to the energy comes from the volume magnetostatic charge given by

$$W_{MS} = \frac{M_s^2 L}{2}\int (m_x \partial_x \Phi + m_y \partial_y \Phi)d^2\vec{r} \quad (15)$$

including the partial derivative of the magnetostatic potential given by

$$\partial_x \Phi = -L\int \frac{(x-x')\nabla \cdot \vec{m}(x',y')dx'dy'}{(\vec{r}-\vec{r}')^2\sqrt{(\vec{r}-\vec{r}')^2 + L^2/4}} \quad (16)$$

where $(\vec{r}-\vec{r})^2 = (x-x')^2 + (y-y')^2$, with a similar expression for $\partial_y \Phi$.

Next use (14) – (16) to obtain the second and fourth order expressions for the magnetostatic energy as a power series in $a/R$. It is remarked that all of the odd contributions will be zero because of the form of the ansatz (12). Moreover, the integral over $\chi$ is trivial for all orders, but before this integral is done it is convenient to make the substitution, $\alpha = \chi' - \chi$. Then the second order contribution can be expressed in dimensionless form by the substitutions, $r = R\xi$ and $r' = R\xi'$ as universal potential integral

$$W_{MS}^{(2)} = \pi M_s^2 \frac{L^2 a^2}{2R} \Phi(\lambda), \quad \lambda = L/R, \tag{17}$$

$$\Phi(\lambda) = \int \frac{(\xi - \xi'^2)(\xi' \cos\alpha - \xi)\cos\alpha}{\Pi\sqrt{\Pi + \lambda^2}} \xi' d\xi' d\xi$$

where the upper limits are $\xi = \xi' = 1$ and to shorten equations the notation $\Pi \equiv \Pi(\xi, \xi', \alpha) = \xi^2 + \xi'^2 - 2\xi\xi' \cos\alpha$ is used. The integral over $\alpha$ can be easily done for the thin disk approximation where the $L^2/4R^2$ contribution in the denominator is neglected giving combinations of elliptic integrals. The remaining radial integrals over $\xi$ and $\xi'$ are done numerically. Similar but more complicated expressions appear for the fourth order term $W_{MS}^{(4)}$ finally giving the magnetostatic energy

$$W_{MS} = \pi M_s^2 \frac{L^2}{R^2}\left(2.83 R a^2 + 60.53 \frac{a^4}{R}\right). \tag{18}$$

The restoring force from (18) can be used with the other forces in (8) to get the vector Thiele equation

$$\left[4\pi\omega - \omega_M \frac{L}{R}\left(2.83 + 121\frac{a^2}{R^2}\right)\right]\vec{e}_r$$

$$+4\pi\left[\sigma I - \alpha\omega \ln(\frac{R}{l_0})\right]\vec{e}_\chi = 0 \tag{19}$$

In the absence of dissipation and current in the linear approximation the first two terms will give the frequency of the gyrotropic mode,

$$\omega = 0.225 \omega_M (L/R), \tag{20}$$

where $\omega_M = 4\pi\gamma M_s$ which is about 30 GHz for permalloy. The dependence on $L/R$ is the same as found in by a different method, with a slight difference in the numerical coefficient [18]. The third term will give the nonlinear frequency shift in a form

$$\omega(a) = \omega\left[1 + 42.76\left(\frac{a}{R}\right)^2\right] \quad (21)$$

The critical current is estimated using the last two terms in (19) with $a = 0$

$$I_c = \frac{\alpha\omega}{\sigma}\ln\left(\frac{R}{l_0}\right), \quad (22)$$

coinciding with the simple estimate based on equations (6, 7).

To estimate the critical current for a typical disk ($R = 200$ nm and $L = 20$ nm) for permalloy ($M_s = 8\times 10^5$ A/m) one obtains $I_c \approx 20 \mu$ A. Larger values of the current $I > I_c$ will result in precession motion of the vortex with amplitude $a$. The dependence of $a$ on the current is determined using the nonlinear term in (21)

$$\frac{a}{R} = 0.153\sqrt{\frac{I - I_c}{I_c}} \quad (23)$$

demonstrating a "soft" non-hysteretic regime of excitation for this mode.

In conclusion, small values of a spin-polarized current can counteract dissipation and result in gyrotropic vortex motion. Since the polarization direction is perpendicular to the vortex plane, this current cannot excite the initial motion, but any small fluctuation of the vortex position or weak and shore in-plane field pulse can produce the initial vortex displacement, which will be amplified by the current. This effect can be observed experimentally by imaging techniques such as time-resolved Kerr microscopy.

**Acknowledgement**

BI was partly supported by INTAS-05-1000008-8112.

**References**

[1] J. C. Slonczewski, J. Magn. Magn. Mater. **159**, L1 (1996).

[2] L. Berger, Phys. Rev. B **54**, 9353 (1996).

[3] S. M. Rezende, F. M. de Aguiar and A. Azevedo, Phys. Rev. Lett. **94**, 037202 (2005).

[4] A. N. Slavin and V. Tiberkevich, Phys. Rev. Lett. **95**, 237201 (2005).

[5] J. Shibata, Y. Nakatani, G. Tatara, H. Kohno and Y. Otani, Phys. Rev. B **73**, 020403(R) (2006).

[6] J.-G. Caputo, Y. Gaididei, F. G. Mertens, and D. D. Sheka Phys. Rev. Lett. **98,** 056604 (2007).

[7] B. A. Ivanov and C. E. Zaspel, Appl. Phys. Lett. **81**, 1261 (2002).

[8] K. Yu. Guslienko, B. A. Ivanov, Y. Otani, H. Shima, V. Novosad, and K. Fukamichi, J. Appl. Phys. **91**, 8037 (2002).

[9] B. A. Ivanov and C. E. Zaspel, Phys. Rev. Lett. **94**, 027205 (2005).

[10] B. A. Ivanov, H. J. Schnitzer, F. G. Mertens and G. M. Wysin, Phys. Rev. B **58**, 8464 (1998).

[11] M. Buess, T. P. J. Knowles, R. Höllinger, T. Haug, U. Krey, D. Weiss, D. Pescia, M. R. Scheinfein, and C. H. Back, Phys. Rev. B **71**, 104415 (2005).

[12] G. Gubbiotti, G. Carlotti, T. Okuno, T. Shinjo, F. Nizzoli, and R. Zivieri, Phys. Rev. B 68, 1844091 (2003).

[13] M. Buess, T. Haug, M.R. Scheinfein and C.H. Back, Phys. Rev. Lett. **94**, 127205 (2005).

[14] K. Yu. Guslienko Appl. Phys. Lett. **89**, 022510 (2006).

[15] A. A. Thiele, Phys. Rev. Lett. **30**, 239 (1973).

[16] D. L. Huber, Phys. Rev. B **26**, 3758 (1982).

[17] A. V. Nikiforov and É. B. Sonin, Sov. Phys. JETP **58**, 373 (1983).

[18] K.Yu. Guslienko, X. F. Han, D. J. Keavney, R. Divan, and S. D. Bader, Phys. Rev. Lett. **96,** 067205 (2006).